\documentclass[journal]{IEEEtran}
\usepackage[caption=false]{subfig}
\usepackage{cite}
\usepackage{amsmath,amssymb,amsfonts}
\usepackage{algorithmic}
\usepackage{graphicx}
\usepackage{textcomp}
\usepackage{xcolor}
\usepackage{gensymb}
\usepackage{booktabs}
\usepackage{soul}

\usepackage[
  separate-uncertainty = true,
  multi-part-units = repeat
]{siunitx}

\usepackage{multirow}
\usepackage{float}

\usepackage{booktabs}

\newcommand{\up}{\mathrm}
\newcommand{\lp}{\left(}
\newcommand{\rp}{\right)}
\newcommand{\ob}[1]{\mkern 1.5mu\overline{\mkern-1.5mu#1\mkern-1.5mu}\mkern 1.5mu}
\newcommand{\mytilde}{\raise.17ex\hbox{$\scriptstyle\mathtt{\sim}$}}

\begin{document}
\bstctlcite{IEEEexample:BSTcontrol}

\author{Ryan~Kaveh*,~\IEEEmembership{Student~Member,~IEEE,}
        Justin~Doong*,~\IEEEmembership{Student~Member,~IEEE,}
        Andy~Zhou,~\IEEEmembership{Student~Member,~IEEE,}
        Carolyn~Schwendeman,
        Karthik~Gopalan,
        Fred~L.~Burghardt,
        Ana~C.~Arias,
        Michel~M.~Maharbiz,~\IEEEmembership{Senior~Member,~IEEE,}
        and~Rikky~Muller,~\IEEEmembership{Senior~Member,~IEEE}

\thanks{*Equally credited authors}
\thanks{R. Kaveh, J Doong, A. Zhou, C. Schwendeman, K. Gopalan, F. Burghardt, and A. C. Arias are with the Department of Electrical Engineering and Computer Sciences, University of California at
Berkeley, Berkeley, CA 94720 USA.}
\thanks{M. M. Maharbiz and R. Muller are with the Department of Electrical
Engineering and Computer Sciences, University of California at Berkeley,
Berkeley, CA 94720 USA, and also with Chan-Zuckerberg Biohub, San Francisco, CA 94158 USA (e-mail: rikky@berkeley.edu).}
}

\title{Wireless User-Generic Ear EEG}
\maketitle

\begin{abstract}
In the past few years it has been demonstrated that electroencephalography (EEG) can be recorded from inside the ear (in-ear EEG). To open the door to low-profile earpieces as wearable brain-computer interfaces (BCIs), this work presents a practical in-ear EEG device based on multiple dry electrodes, a user-generic design, and a lightweight wireless interface for streaming data and device programming. The earpiece is designed for improved ear canal contact across a wide population of users and is fabricated in a low-cost and scalable manufacturing process based on standard techniques such as vacuum forming, plasma-treatment, and spray coating. A 2.5~$\times$~2.5~cm\textsuperscript{2} wireless recording module is designed to record and stream data wirelessly to a host computer. Performance was evaluated on three human subjects over three months and compared with clinical-grade wet scalp EEG recordings. Recordings of spontaneous and evoked physiological signals, eye-blinks, alpha rhythm, and the auditory steady-state response (ASSR), are presented. This is the first wireless in-ear EEG to our knowledge to incorporate a dry multielectrode, user-generic design.  The user-generic ear EEG recorded a mean alpha modulation of 2.17, outperforming the state-of-the-art in dry electrode in-ear EEG systems.
\end{abstract}


\IEEEpeerreviewmaketitle

\section{Introduction}
\IEEEPARstart{B}{rain} computer interfaces (BCI) enable users to communicate with and control the growing number of available computing, sensing, and actuating tools. A completely non-invasive technique, electroencephalography (EEG), has become increasingly established as a safe clinical and research method to record the brain’s electrical activity from the scalp surface \cite{Teplan2002}. Clinically, EEG is used to monitor neurological disorders related to epilepsy, sleep, and strokes\cite{Noachtar2009}\cite{Wu2016}. In research settings, EEG has been used to study neuroscience such as cortical networks and memory formation \cite{sauseng2005}. EEG has been further extended to BCIs, where compact (\textless 5 channels) systems have recorded steady-state auditory and visually evoked potentials, low-frequency neural waves (e.g., alpha rhythm), and auditory and visual event-related potentials (ERPs) (e.g., P300)  to perform choice selection, cursor movement, and prosthetic control \cite{Schalk2004}\cite{Duvinage2012}\cite{Muller-putz2008}.

\begin{figure}[tb]
\centerline{\includegraphics[width=2.75in]{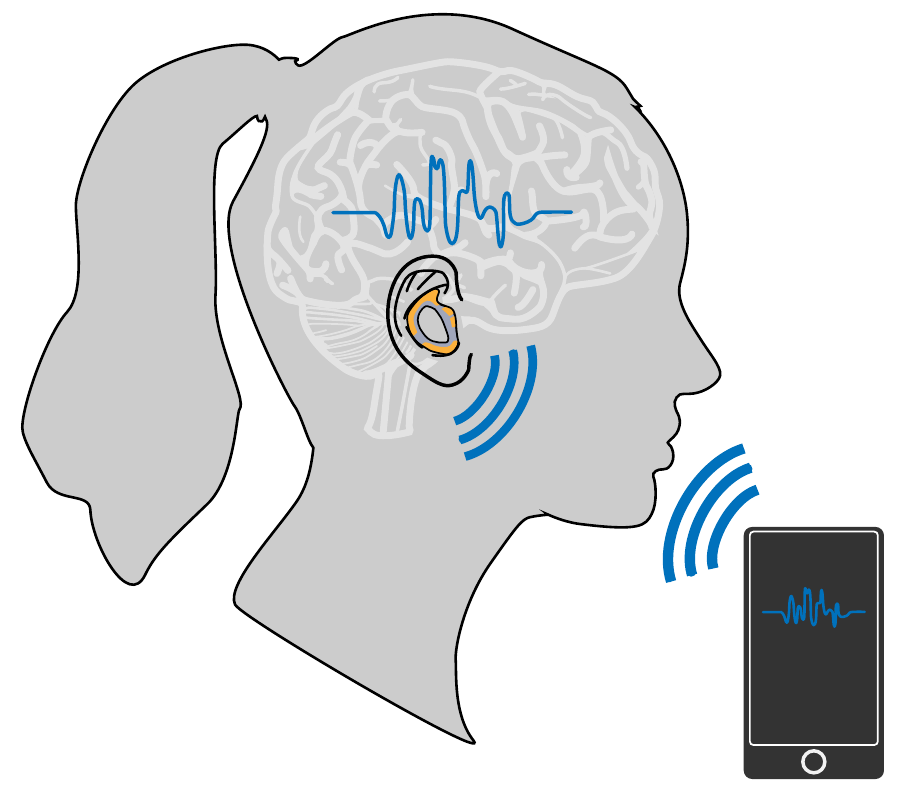}}
\vspace{-8pt}
\caption{Envisioned BCI discreetly connecting user to devices.}
\label{prop_system}
\vspace{-18pt}
\end{figure}

Current clinical setups have shortcomings that preclude them from directly translating to consumer contexts. Clinical EEG systems rely on wet electrodes (i.e., electrodes with hydrogel) to make electrode-skin contact through hair, reduce the electrode-skin impedance (ESI), and provide mechanical stability (hydrogels act as adhesives). These hydrogels dry out over time and result in increased interference susceptibility and signal-to-noise ratio (SNR) degradation \cite{Searle2000}. Moreover, the electrode application process requires skin-abrasion on every electrode-site, and results in hair-loss or skin lesions \cite{Joellan2014}. The process of individually applying the electrodes is also time-consuming (especially on subjects with long hair) and must be performed by a trained technician\cite{Teplan2002}. In addition, clinical setups use long wires to connect electrodes to a recording module, which exacerbates interference and motion artifacts while impairing subject movement \cite{Tatum2011}. Advancements in clinical EEG systems (e.g., wireless recorders, mesh vests to hold wires, and caps to hold electrodes) have improved usability for inpatient care, but they remain prohibitively bulky for day-to-day use outside the laboratory.

\begin{figure*}[htb]
    \centering
    \includegraphics[width=6.25in]{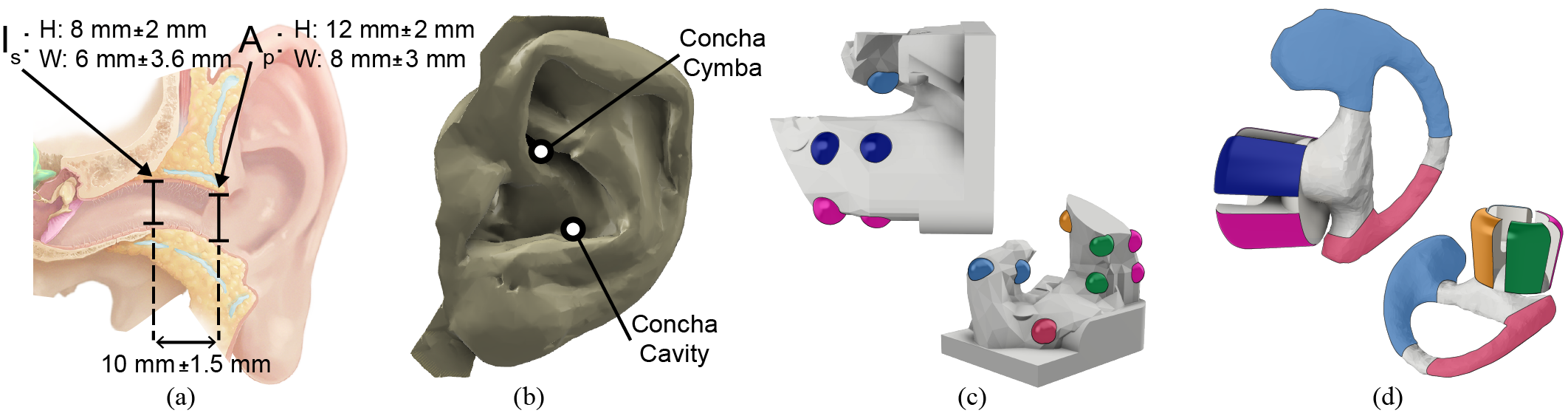}
    \vspace{-8pt}
    \caption{(a) Ear canal with average measurements from \cite{Staab2000} \cite{ear}. (b) High-resolution subject ear scan. (c) Earpiece custom-fit for (b) shown from two angles. Correlated electrode pairs are highlighted by the same color. (d) User-generic earpiece. Electrodes highlighted with corresponding group colors from (c).}
    \label{earpiece_prog}
    \vspace{-17pt}
\end{figure*}

Recent developments in dry electrodes and recording electronics integration attempt to address these shortcomings. Dry electrodes eliminate the use of hydrogel, thus simplifying the electrode application process. This improved usability results in higher impedance relative to wet electrodes \cite{Kappel2019}. Higher impedance not only results in greater interference susceptibility but also greater electrode-related noise. To lower ESI in dry electrodes, the state-of-the-art has employed microneedles, electrode fingers, conductive composites, and nanowires \cite{Lopez-Gordo2014}. However, these head-worn dry electrodes are either uncomfortably large or require skin preparation. Prolonged use of these devices often results in skin irritation or lesion formation, thus limiting their use \cite{tasca2019}. Commercial companies have incorporated electrodes and recording electronics designed to accommodate dry electrode properties into a wireless headset. While more compact than clinical systems, these headsets still require tedious training, electrode preparation, and cover large parts of the scalp. Furthermore, they lack the required comfort, discreetness, and motion-artifact/interference robustness for daily, public use \cite{XU2018}.

To enable comfortable BCIs that can be integrated into commonly used devices and intuitively used, sensing EEG from inside the ear (in-ear EEG) has been proposed \cite{Kidmose2013}\cite{Looney2012}\cite{Goverdovsky2016}\cite{Gert2019}. Recent work has demonstrated that both steady-state and transient evoked potentials (30--80~Hz) can be recorded with both sensing and referencing electrodes placed in-ear. Moreover, low frequency neural rhythms (1--30~Hz) and electrooculography (EOG) signals (eye blinks) have been recorded \cite{Zibrandtsen2017}. While this close-proximity electrode scheme results in less spatial coverage and smaller amplitude signals relative to scalp EEG systems \cite{Looney2012}, existing demonstrations have exhibited promising results for discreet BCIs that could enable new ways of interacting with everyday devices (Fig.~\ref{prop_system}).

Mirroring clinical EEG design progression, initial demonstrations of in-ear EEG used user-specific molded earpieces equipped with wet Ag electrodes \cite{Kidmose2013}\cite{Looney2012}\cite{Zibrandtsen2017}. These customized earpieces guaranteed consistent electrode-skin contact while the wet electrode gel (and the accompanying skin preparation) reduced ESI. Though these systems were successful first steps, a widely adopted BCI cannot rely on users to perform skin preparation and gel application due to the associated skin damage \cite{Joellan2014}. Recent works have incorporated small, dry electrodes in custom earpieces to improve usability at the cost of increased ESI, noise, and signal-quality degradation. Regardless, while both wet and dry electrode versions have successfully recorded various spontaneous and evoked EEG signals, custom molded earpieces are not conducive to low-cost, large-scale manufacturing and user adoption. Rather, a low-cost, user-generic solution is required. 

To make in-ear EEG widely usable, this paper presents a wireless in-ear EEG recording platform (user-generic ear EEG). To fit a large number and wide range of users, the design is based on dry electrodes in a user-generic rather than individualized design. To achieve scalability and low-cost, earpiece manufacturing avoids non-standard materials (carbon nanotubes, graphene, etc.) and makes use of low-cost processes. To enable comfortable, ambulatory use, low-noise recording electronics are integrated in a compact, wireless module. The electrodes were characterized with ESI and electrode dc offset (EDO) measurements. The wireless module was characterised for signal to noise ratio (SNR), interference robustness, and battery life. Finally, a user study was performed to measure three classes of electrophysiological signals across three users. Physiological measurements of eye blinks, alpha modulation, and auditory steady-state response (ASSR) are presented and compared with wet scalp EEG recordings and state-of-the-art in-ear EEG.

The manuscript is organized as follows: Section~II describes the user-generic, dry-electrode earpiece design, fabrication, and characterization. Section~III motivates in-ear EEG recording requirements and introduces the wireless neural recording module. Section~IV presents the system setup, experimental designs, and physiological measurement results. Section~V provides a summary and comparison with the state-of-the-art.

\section{User-Generic, Dry-Electrode Earpiece}

User-generic in-ear EEG designs replace customized earpieces with compliant electrodes that can apply pressure on the skin to improve mechanical stability and reduce ESI. This spring compression can be achieved through the use of conductive rubbers, foams, or stiff substrates with cantilever geometries. While prior demonstrations of dry conductive composite earpieces are comfortable, they require large area electrodes due to their low conductivity. These examples also rely on wet-electrode references placed outside the ear and are limited to a single channel \cite{HoonLee2014}\cite{Dong2016}. Metal electrodes on compliant substrates could provide comparable comfort and mechanical stability to conductive composites while also enabling multielectrode designs and the use of low-cost materials.

\begin{figure}[tbp]
\centerline{\includegraphics[width=3in]{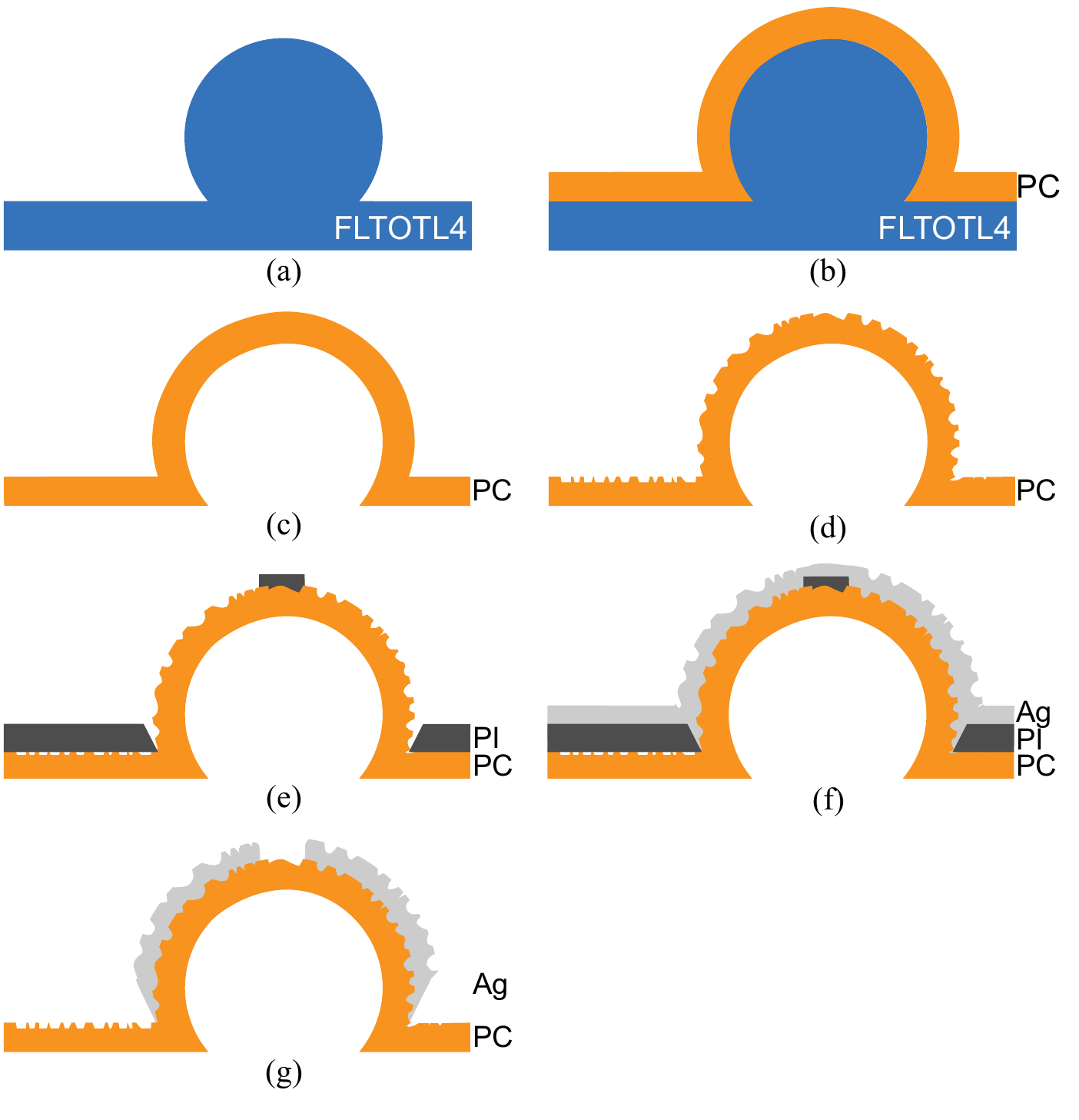}}
\vspace{-8pt}
\caption{Manufacturing process diagram cross section. (a) 3D-printed master mold. (b) Polycarbonate (PC) thermoformed around mold. (c) Hollow PC liberated from master. (d) Sanded PC after alcohol bath. (e) Nitrogen plasma treated earpiece with polyimide (PI) mask. (f) Ag spray-coated PC. (g) PC with independent electrodes.}
\label{manuf_proc}
\vspace{-16pt}
\end{figure}

\subsection{Electrode and Earpiece Design}

 To find unique recording sites in the ear, preliminary measurements were taken with an initial user-specific earpiece designed to provide consistent electrode-skin contact. The initial earpiece was designed based on a high-resolution scan of a subject’s ear (Fig.~\ref{earpiece_prog}(b)). Ten 12~mm\textsuperscript{2} electrodes were spaced evenly across the earpiece’s surface to record along much of the outer ear (Fig.~\ref{earpiece_prog}(c)). Principle electrode locations for the user-generic design were selected from these initial ten sites based on correlations in physiological measurements. Eye blinks and alpha band modulations were recorded in a subject with the customized earpiece (see Section~IV for signal background). Pearson correlation coefficients were calculated for each electrode pair from four eye blink measurement trials and four alpha band modulation experiments. Average correlation coefficients for each electrode pair were then used as criteria for combining electrodes into a larger electrode (to reduce ESI). Electrode pairs that were \textgreater70\% linearly correlated were grouped together (highlighted by the same colors in Fig.~\ref{earpiece_prog}(c)). Electrodes with the lowest degree of correlation with all other electrodes (\textless20\%) were selected as referencing locations. The final design includes four in-ear electrodes (for sensing) and two out-ear electrodes (for referencing). Consolidating ten small electrodes into six larger electrodes reduces ESI, noise, and interference, and promotes increased electrode-skin contact across different ear canal shapes. Furthermore, this electrode arrangement delivers greater flexibility with two large electrodes that can serve as reference on the outer ear. Due to user-to-user variation in ear morphology, the concha cymba may perform as a better reference location for an individual than the concha cavity (or vice versa).

The earpiece structure was further informed by anatomical measurements of the average ear canal. The three most important ear dimensions (the aperture, isthmus, and length highlighted in Fig.~\ref{earpiece_prog}(a)) are reported to have a normal distribution across large human populations \cite{Staab2000}. Thus, three generic earpiece candidates were designed with these average measurements and then scaled up and down by a standard deviation to make small and large earpieces. Each design was tested by multiple individuals for comfort and physical stability during three activities: walking, running, and jumping. The most comfortable user-reported design that remained secure in the ear through all movements was selected as the final structure (Fig.~\ref{earpiece_prog}(d)). The selected earpiece leverages four outward cantilevers to apply independent pressure on each in-ear electrode. This compliance coupled with the enlarged electrode area minimizes ESI across multiple users relative to the custom fitted earpiece and other dry electrode-based earpieces \cite{Kappel2019}. The final design has four 60~mm\textsuperscript{2} electrodes around the ear canal's aperture and two larger, 4~cm\textsuperscript{2} electrodes along the outer ear's concha cymba and concha cavity (Fig.~\ref{earpiece_prog}(a)~\&~(b)).

\begin{figure}[tbp]
\centerline{\includegraphics[width=3.25in]{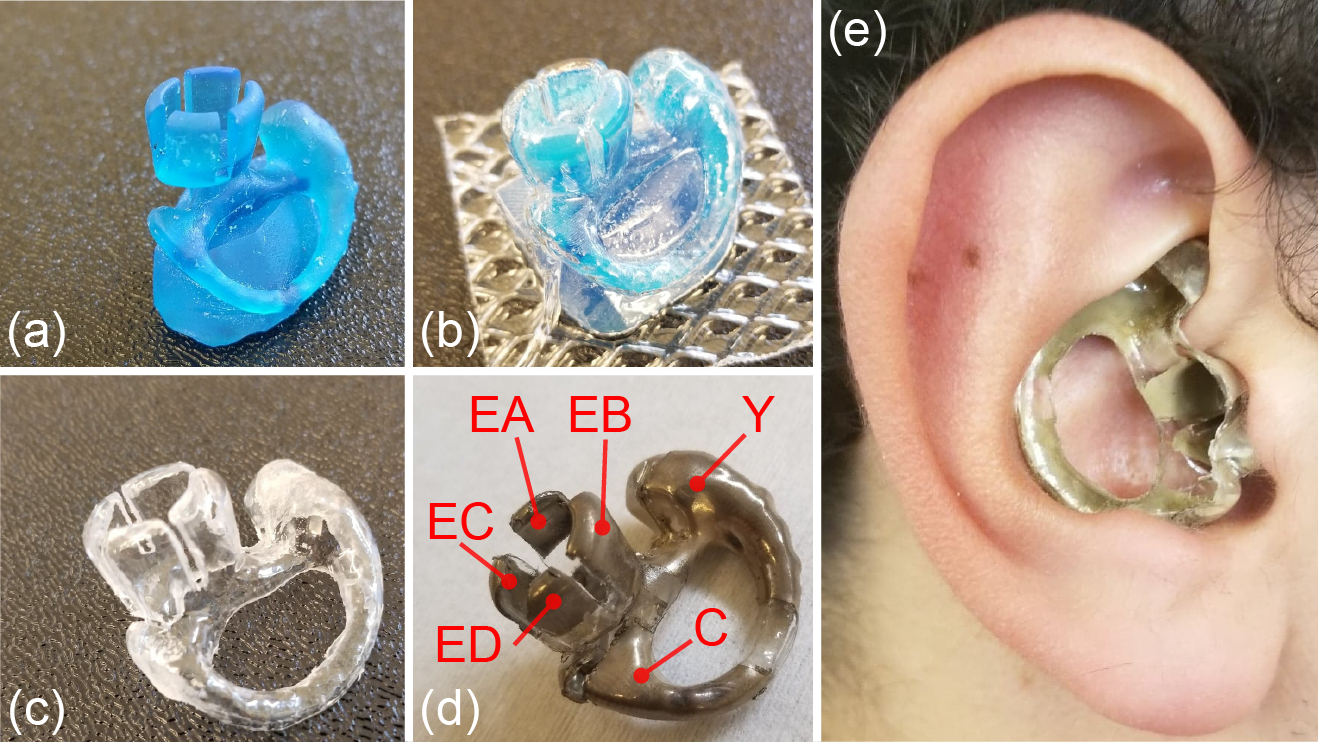}}
\vspace{-5pt}
\caption{Earpiece construction and in-ear fit. (a) 3D printed master mold. (b) Thermoformed earpiece. (c) Liberated earpiece. (d) Spray-coated earpiece. (e) Earpiece in-ear fit.}
\label{manuf_photos}
\vspace{-16pt}
\end{figure}

\begin{figure*}[tbp]
\centerline{\includegraphics[width=6in]{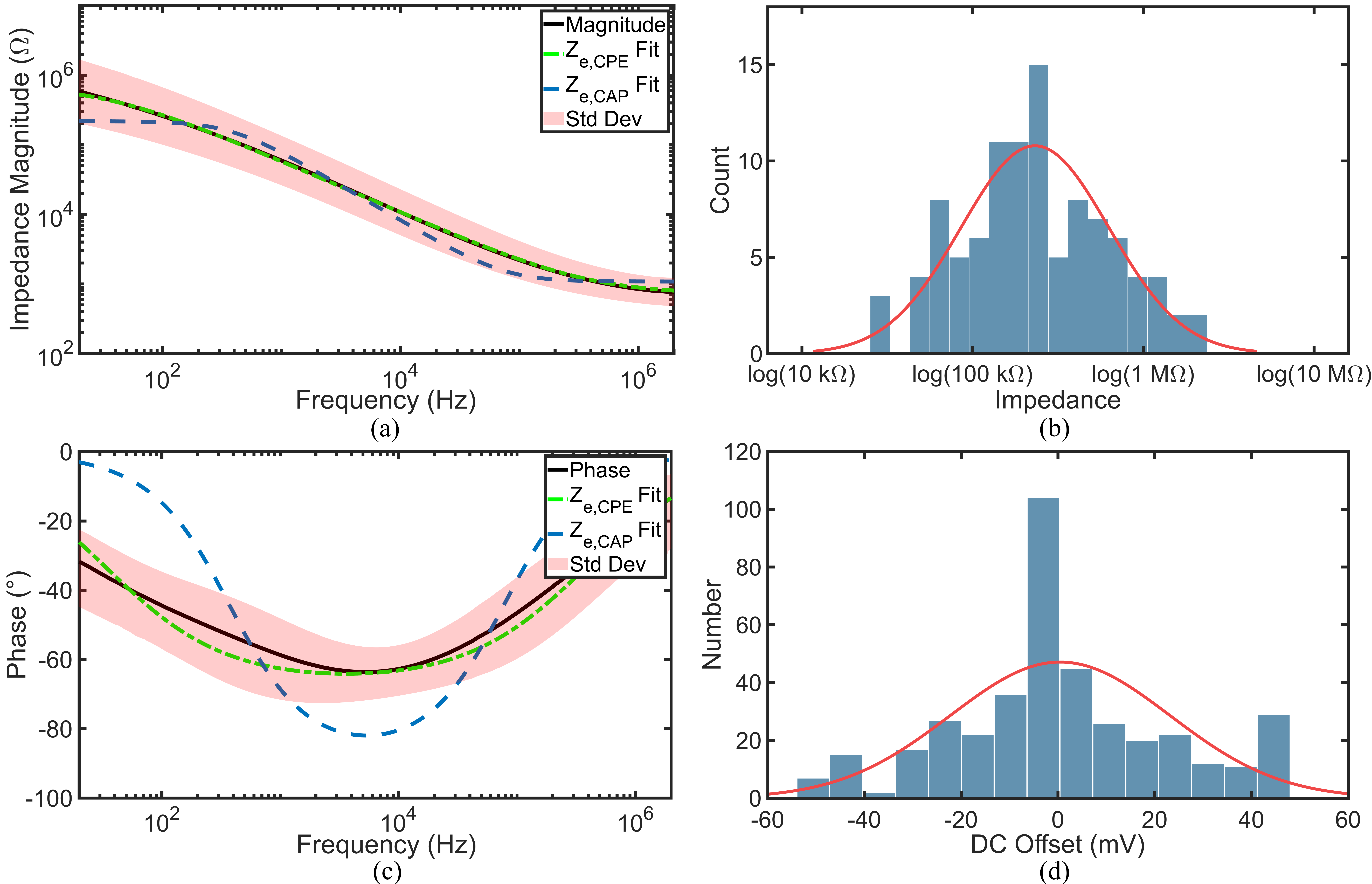}}
\vspace{-10pt}
\caption{Electrical characterizations of in-ear Ag electrodes across three users. (a) Average magnitude of electrodes with fitted $Z_\up{e,CPE}$ and $Z_\up{e,CAP}$ models (n=101). (b)  Histogram plotting the 50~Hz impedance of every impedance spectroscopy trial (n=101). (c) Average electrode skin impedance (ESI) phase with fitted electrode models (n=101). (d) Histogram plotting electrode DC offset (EDO) measurements across three users (n=395).}
\label{elec_char}
\vspace{-10pt}
\end{figure*}

\vspace{-10pt}
\subsection{Earpiece Fabrication}

To build the user-generic earpieces, a low-cost, repeatable manufacturing process based on thermoforming was developed (A sample process diagram is shown in Fig.~\ref{manuf_proc}). This thermoforming-based process enables the rapid molding of multiple earpieces in a single step with minimal tooling. Earpiece construction begins by 3D printing a heat-resistant master mold with an ultraviolet cured polymer resin (Formlabs, Tough resin FTOTL4) (Fig.~\ref{manuf_proc}(a) \& Fig.~\ref{manuf_photos}(a)). Next, a 30~mil sheet of polycarbonate, selected for its flexibility and low-reactivity with isopropyl alcohol (IPA), is heated past its glass transition temperature and thermoformed around the master mold using a vacuum form chamber (450DT, Formech) (Fig.~\ref{manuf_proc}(b) \& Fig. \ref{manuf_photos}(b)). The master mold is then removed from the polycarbonate (Fig.~\ref{manuf_proc}(c) \& Fig.~\ref{manuf_photos}(c)), leaving a hollow earpiece that is compliant and provides space for wire routing. After thermoforming, the earpiece is sanded (Fig.~\ref{manuf_proc}(d)) to increase the effective surface area of the substrate, cleaned with compressed air, and then placed in an IPA bath. A laser cut polyimide (PI) mask is then applied to the earpiece (Fig.~\ref{manuf_proc}(e)) which is then plasma treated with nitrogen to increase the substrate’s surface energy and improve surface adhesion. The treated, masked earpiece is spray-coated with Ag to deposit 15 \SI{}{\micro m} thick electrodes (Fig.~\ref{manuf_proc}(f)). Finally, the PI mask is removed (Fig.~\ref{manuf_proc}(g) \& Fig.~\ref{manuf_photos}(d)), and wires are cold-soldered to each electrode with a heat-cured silver epoxy (Epo-tek, H20E-D). Each silver epoxy bump is passivated with an ultraviolet cured epoxy. The final earpieces can be easily cleaned with IPA and re-used without loss of structural integrity. This process can be easily repeated through reuse of the heat resistant master mold and quickly adapted to different designs, thermoplastics, electrode shapes, and materials. Acrylic, for its rigidity, may be used for the substrate while other metals (e.g., Au for longevity) can be electroplated on the Ag base layer. This process is improved over \cite{Kaveh2019} through the addition of new surface treatment steps, the surface roughening and plasma treatment (Fig.~\ref{manuf_proc}(d)~\&~(e)), that result in an increased effective electrode area (and thus reduced ESI) and electrode lifetime.

To measure the in-ear electrodes’ (EA, EB, EC, and ED) compliance, each electrode was strained with a force gauge. The average electrode spring constant was 171~N/m with a standard deviation of 5~N/m (n=50). This inherent spring allows each electrode to apply approximately 90~kPa (at 50\%~strain). This pressure is on the order of polyurethane foam (at 50\%~strain) \cite{Rusch1969}.

\subsection{Electrode Electrical Characterization}
The electrodes were characterized by  impedance spectroscopy, measurements of EDO, and electrode skin interface noise analysis. ESI and EDO characterizations were performed on all three subjects with the earpiece inside the ear without any skin cleaning or preparation to simulate realistic, day-to-day use.

\subsubsection{Electrode Skin Impedance}\label{ESI}
\begin{figure}[tbp]
\centerline{\includegraphics[width=2in]{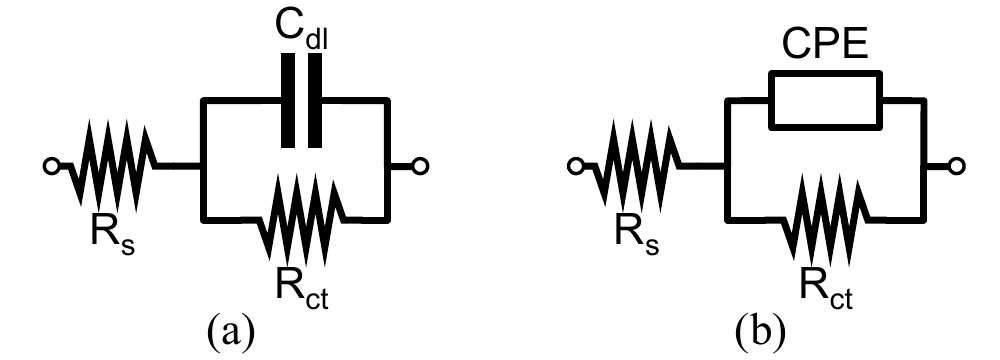}}
\vspace{-10pt}
\caption{(a) Capacitive electrode model $Z_\up{e,CAP}$. (b) Constant phase element electrode model $Z_\up{e,CPE}$.}
\label{elec_models}
\vspace{-10pt}
\end{figure}

All ESI measurements were performed between each dry electrode on the user-generic earpiece and a wet electrode placed on the subject's ipsilateral mastoid. No skin preparation was performed before each trial, and measurements sessions were repeated over the course of six months. Since wet electrodes have an order of magnitude lower impedance than dry electrodes of the same size \cite{Searle2000}, the impedance measurements are dominated by the single dry ESI. All measurements were performed with an LCR meter (E4980A, Keysight) and results were fitted to two equivalent circuit models (spectra shown in Fig.~\ref{elec_char}(a)~\&~(b), circuit models shown in Fig.~\ref{elec_models}). One model comprises resistors and a constant phase element (CPE) and the other comprises resistors and a capacitor. A CPE is an equivalent electrical model for a double layer and is often used as a measure of the electrode-skin interface's non-faradaic impedance. It is modeled by

\begin{table}[t]
\setlength{\tabcolsep}{6pt}
\caption{Electrode model fit parameters}
\begin{center}
\begin{tabular}{rccccc}\toprule
Model & $R_\up{s} (\Omega)$ & $R_\up{ct} (\Omega)$ & $C_\up{dl} (F)$   & $Q_\up{CPE}$ & $n_\up{CPE}$\\ \midrule
$Z_\up{e,CAP}$ & 1.08 k & 217 k & 1.94 n & -- & --\\
$Z_\up{e,CPE}$ & 704  & 793 k & -- & 0.793 x$10^{-9}$ & 0.760\\ \bottomrule
\end{tabular}
\end{center}
\vspace{-20pt}
\label{tb:model_params}
\end{table}

\begin{equation}
Z_\up{CPE} = \frac{1}{\lp\jmath\omega\rp^n Q},
\end{equation}
where $0 < n \leqslant 1$. $Q$ is a measure of the magnitude of $Z_\up{CPE}$ while $n$ fits the bilayer phase offset. The CPE based electrode model's impedance can be described by
\begin{equation}
Z_\up{e,CPE} = R_\up{s} + \frac{R_\up{ct}}{1 + \lp \jmath\omega\rp^n Q R_\up{ct}},
\end{equation}
while the total impedance of the capacitive electrode model is given by
\begin{equation}
Z_\up{e,CAP} = R_\up{s} + \frac{R_\up{ct}}{1 + \jmath\omega  C_\up{dl} R_\up{ct}},
\end{equation}

\noindent where $R_\up{s}$ is spread resistance, $R_\up{ct}$ is the charge-transfer resistance, and $C_\up{dl}$ is the double layer capacitance. While both models are commonly used for electrodes \cite{Franks2005}, $Z_\up{e,CPE}$ fits the average impedance spectrum better than $Z_\up{e,CAP}$ since CPEs can model phase shifts less than \ang{90} (Fig.~\ref{elec_char}(a)~\&~(c)). The impedance measurements can be modelled using a log normal distribution described by  equation (4), where the mean value \(\mu\) is 231~k\(\Omega\) and the standard deviation factor \(\sigma\) is 2.711 (Fig.~\ref{elec_char}(b)).
\begin{equation}
X = 10^{\mu + \sigma V}
\end{equation}
No major degradation was observed in the ESI despite heavy re-use and cleaning over the course of six months nor did measured impedance values increase over time. Lastly, ESI was also used to grade electrode-skin contact. Acceptable contact was determined by a 50~Hz ESI \textless 1~M\(\Omega\). 90\% of all measurements meet this criterion. At 50~Hz, the interface has an average impedance of 392~k\(\Omega\) and phase of -39°.

\subsubsection{Electrode DC Offset}
The electrode dc offset (EDO) of the electrodes relative to the reference has implications on the required neural recording input range  or offset cancellation range. Maintaining a low EDO is desirable to keep the recording frontend in its linear range and out of saturation. Measurements were taken between each dry sensing electrode on the user-generic earpiece and the dry reference electrode (placed on the concha cymba) with the WANDmini recording frontend (see Section~III) supporting a 400~mV input range. No skin cleaning or abrasion was performed. Fig. \ref{elec_char}(d) reports final values after 7 seconds for 395 measurements taken from the three subjects. The EDO measurements are normally distributed, with a mean and standard deviation of 0.444~mV and 22.7~mV respectively. The minimum and maximum EDO values are respectively -49.6~mV and 45.1~mV.

\begin{figure}[tp]
\centerline{\includegraphics[width=3in]{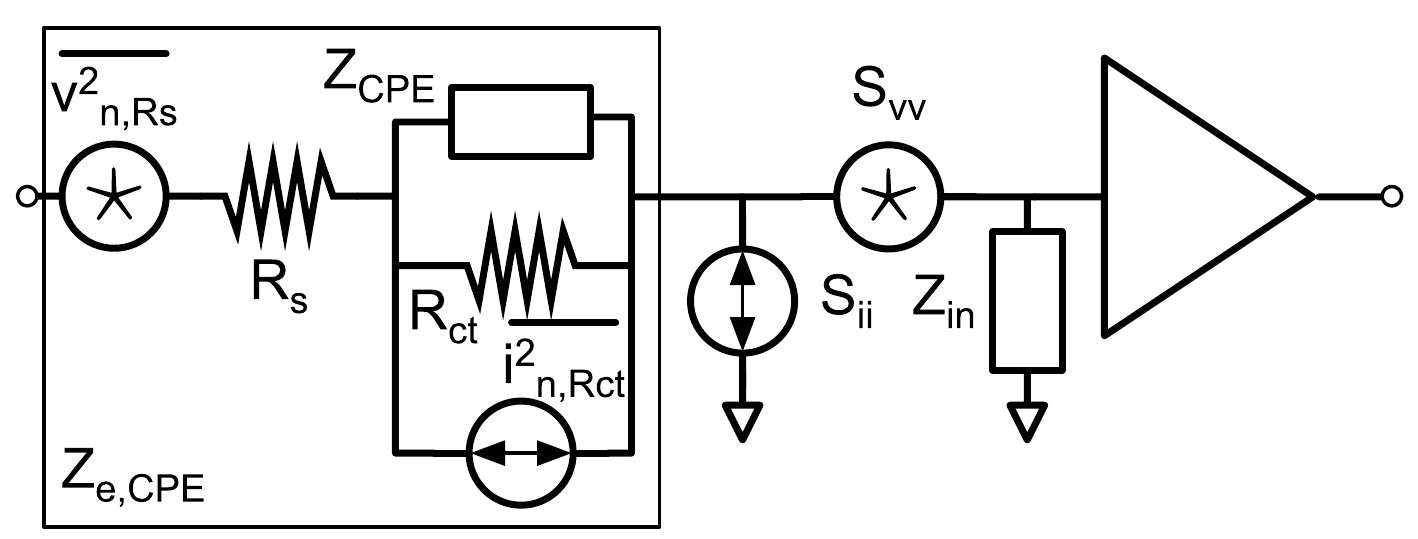}}
\caption{Noise sources at input of readout frontend: electrode noise $v_{\up{n,}R_\up{s}}$ and $i_{\up{n,}R_\up{ct}}$, input-referred frontend voltage noise $v_{\up{n,a}}$, and input-referred frontend current noise $i_{\up{n,a}}$.}
\label{noise_ckt}
\vspace{-20pt}
\end{figure}

\subsubsection{System Noise}\label{noise_math_sec}
To robustly record in-ear EEG, the total system noise floor must be low enough to provide sufficiently high SNR for all features of interest across a wide population of users. State of the art typically quantifies the noise arising from the the electrode and the recording electronics independently. However, to analyze the total input noise, all components and their interactions must be considered, as shown in Fig.~\ref{noise_ckt}. The electrode noise can be modeled as the thermal noise contributions of $R_\up{s}$ ($\ob{v_{\up{n,}R_\up{s}}^2}$) and $R_\up{ct}$ ($\ob{i_{\up{n,}R_\up{ct}}^2}$) \cite{Chi2010}. Typically, noise from recording electronics is modeled as a combination of uncorrelated input-referred voltage and current noise. However, these two noise sources can be correlated, since they may be generated by the same devices \cite{Razavi}. In this case, the noise can be modeled as a combination of the input-referred voltage noise $v_{\up{n,a}}$ and current noise $i_{\up{n,a}}$ as well as a correlated noise contribution \cite{Xu2000}. $S_\up{vv}$ and $S_\up{ii}$ are the power spectral densities (PSD) of the voltage and current noise respectively while $S_\up{vi}$ describes the PSD of the correlated component of the noise. The total noise $S_\up{nn}$ referred to the input of the electrode can be described by
\begin{align}
\begin{split}
    S_\up{nn} ={} &\ob{v_{\up{n,}R_\up{s}}^2} \\
    &+ \ob{i_{\up{n,}R_\up{ct}}^2} \left|\frac{R_\up{ct} || Z_\up{CPE}}{\lp R_\up{ct} || Z_\up{CPE}\rp + R_\up{s} + Z_\up{in}} \cdot \lp Z_\up{in} + Z_\up{e,CPE}\rp \right|^2 \\
    &+ S_\up{vv}
    + S_\up{ii} \left|Z_\up{e,CPE}\right|^2 \\
    &+ 2 \up{Re}\left\{ S_\up{vi} \frac{Z_\up{in}Z_\up{e,CPE}}{Z_\up{in} + Z_\up{e,CPE}} \right\}.
\end{split}
\end{align}
Due to the small value of $R_\up{s}$ ($\leqslant$1~k$\Omega$) relative to $R_\up{ct}$ and $Z_\up{e,CPE}$ (both $>$100~k$\Omega$), the contribution of the first term is negligible. The contribution of $S_\up{ii}$ and $S_\up{vi}$ are often ignored since the typical wet electrode impedance is sufficiently small such that other noise sources dominate \cite{Chi2010}\cite{Xu2013}. However, in the context of dry electrode recording it cannot be ignored due to the high electrode impedance, therefore the current noise contribution can increase the noise floor and degrade SNR. Section~\ref{WANDmini} and Fig.~\ref{noise} detail the measured electrode noise contributions to the user-generic ear EEG system noise.

\section{Wireless Neural Recording Module}\label{WANDmini}

\begin{figure}[tp]
\centerline{\includegraphics[width=3in]{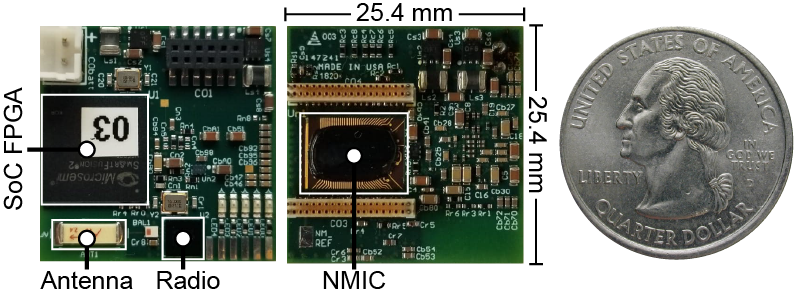}}
\caption{WANDmini neural recording module beside scale quarter.}
\label{wand_mini_pcb}
\end{figure}

\begin{figure}[tp]
\centerline{\includegraphics[width=3.25in]{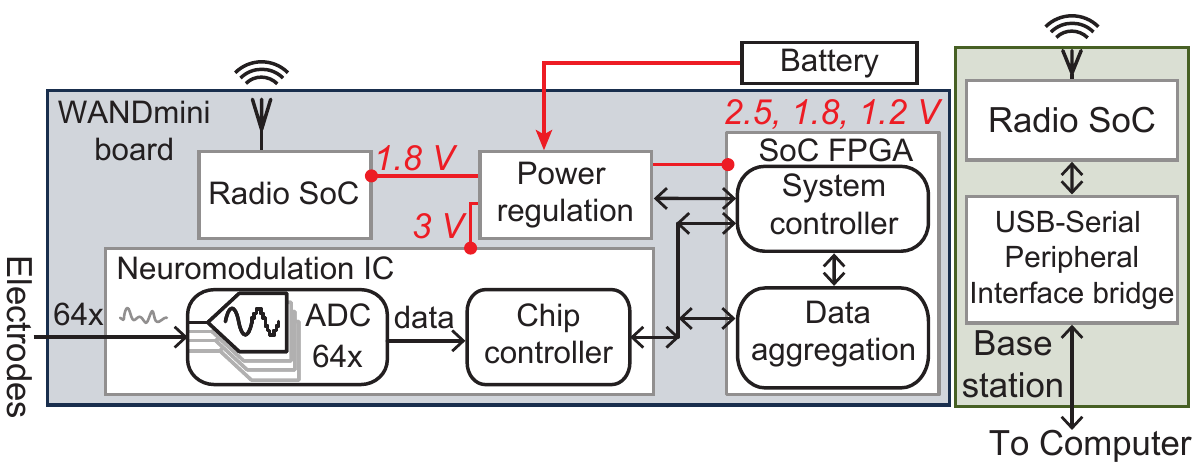}}
\caption{WANDmini system architecture as used in the user-generic ear EEG.}
\label{wand_mini_system}
\vspace{-16pt}
\end{figure}

Recording Ear EEG from a compact, wireless recording module enables ambulatory neural recording, improves ease of use, and minimizes motion artifacts. In this work, Ear EEG signals are acquired using a miniature, wireless, artifact-free neuromodulation device (WANDmini), a low-profile, custom neural recording system that streams recorded data over Bluetooth Low Energy (BLE) to a base station connected to a laptop. WANDmini is derived from a previous design for a wireless, artifact-free neuromodulation device (WAND) \cite{Zhou2019}, reduced to form factor of 2.5~$\times$~2.5~cm\textsuperscript{2}, and embedded with custom firmware (Fig. \ref{wand_mini_pcb}). Recording and digitization are performed by a custom neuromodulation IC \cite{Johnson2017} (NMIC, Cortera Neurotechnologies, Inc.) integrated with 64 digitizing frontends, thereby expandable to recording applications with higher electrode counts. NMIC and WANDmini specifications are listed in Table~\ref{tb:wand}, and the WANDmini system block diagram is shown in Fig.~\ref{wand_mini_system}.

\begin{figure}[tp]
\centerline{\includegraphics[width=3.5in]{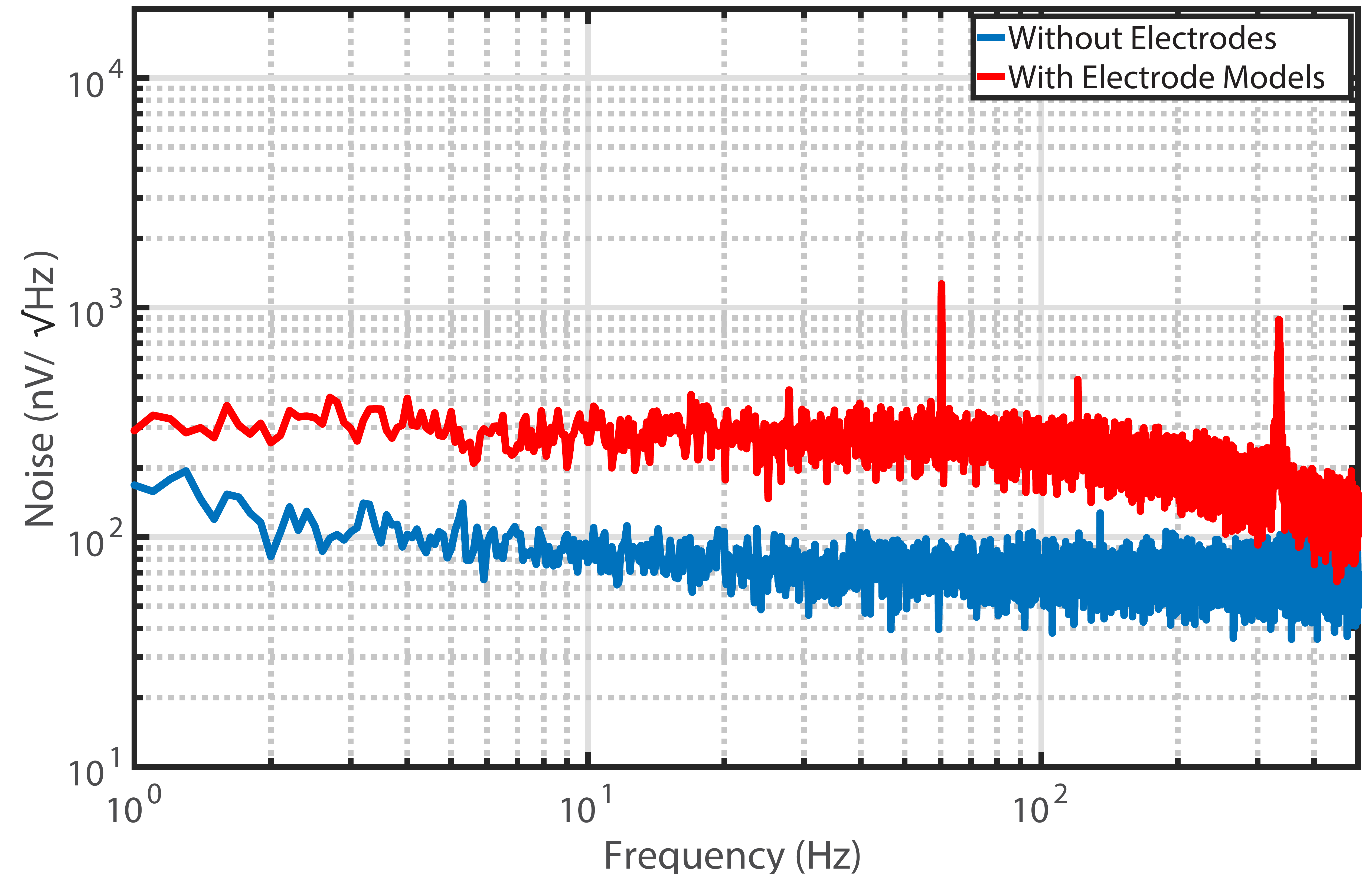}}
\caption{Measured input-referred noise spectrum with and without two capacitive electrode models $Z_\up{e,CAP}$ at the amplifier input.}
\label{noise}
\vspace{-15pt}
\end{figure}

The NMIC was selected for its low power and high dynamic range, supporting a 100--400~mV input range with a flat input-referred noise voltage spectrum of 70~nV/$\sqrt{\text{Hz}}$, which is lower than the electrode thermal noise voltage (Fig.~\ref{noise}). The analog-to-digital converters (ADCs) have a resolution of 15~bits and sample at 1~kS/s, providing sufficient resolution and bandwidth for EEG signals. The wide linear input range can accommodate the large electrode dc offsets (Fig.~\ref{elec_char}(d)), and provides robustness to interference. The total harmonic distortion (THD) with a full-scale input is 0.012\%, maintaining linearity in the presence of large interferers \cite{Zhou2019}. The input range is expandable to 400~mVpp to accommodate larger electrode offsets at the expense of quantization noise. The front-ends achieve this large range with a mixed-signal architecture that includes the ADC into the feedback loop to reduce the required gain and signal swings. Many systems handle EDO by simply AC coupling, but recording EDO provides additional information on electrode performance, and eliminates the long settling times associated with AC coupling capacitors. The NMIC also has stimulation and impedance measurement capabilities, which are not used in this study.

\begin{table}[b]
\vspace{-20pt}
\setlength{\tabcolsep}{17pt}
\caption{NMIC and WANDmini Specifications}
\begin{center}
\begin{tabular}{rc}
\toprule 
\multicolumn{2}{c}{\textbf{NMIC}}                   \\ \midrule
Max Recording Channels       & 64                     \\ 
Recording Channels Used      & 5                      \\ 
Input Range                  & 100 mVpp                \\ 
Input-Ref Voltage Noise      & 70 nV/$\sqrt{\text{Hz}}$ \\ 
Input-Ref Current Noise      & 286 fA/$\sqrt{\text{Hz}}$\\ 
THD                          & 0.012\% (100 mVpp input) \\ 
ADC Resolution               & 15 bits                \\ 
ADC Sample Rate              & 1 kS/s                 \\ 
Power                        & 700 \SI{}{\micro W}    \\ \midrule 
\multicolumn{2}{c}{\textbf{WANDmini}}               \\ \midrule 
Wireless Data Rate           & 2 Mbps                 \\ 
Board Dimensions               & 25.4 mm $\times$ 25.4 mm\\ 
Weight (w/o battery)         & 3.8 g                  \\ 
Supply Voltage               & 3.5 V                    \\ 
Power                        & 46 mW                    \\ 
Battery Life                & \mytilde44 Hours    \\ \bottomrule
\end{tabular}
\label{tb:wand}
\end{center}
\vspace{-10pt}
\end{table}

The measured input-referred noise spectrum is shown in Fig.~\ref{noise} with and without two capacitive electrode models representing the electrodes at the amplifier input. The electrodes were modeled as $Z_\up{e,CAP}$ with parameters $R_\up{s} = \SI{1}{k\ohm}$, $R_\up{ct} = \SI{270}{k\ohm}$, and $C_\up{dl} = \SI{2.4}{nF}$ (see section~\ref{ESI}). As described in Section~\ref{noise_math_sec}, the input current noise contributes to the overall system noise when coupled to a high impedance electrode. An input current noise of 286~fA/$\sqrt{\text{Hz}}$ and a correlated noise of 11.2~$\sqrt{\text{W/Hz}}$ were measured for the NMIC recording channel. Without electrodes present, a noise spectrum of 70~nV/$\sqrt{\text{Hz}}$ and an integrated noise (1--500~Hz) of 1.6~\SI{}{\micro V}rms were measured. With the electrodes present, the low frequency noise spectrum increased to \mytilde270~nV/$\sqrt{\text{Hz}}$ and resulted in an integrated noise of 4.65~\SI{}{\micro V}rms, sufficient to record ear EEG with SNR $>$15~dB. Since the electrodes contribute \mytilde90~nV/$\sqrt{\text{Hz}}$, a custom design with lower current noise terms could decrease the total integrated noise by up to 2~\SI{}{\micro V}rms.

WANDmini system control and data aggregation are performed by an on-board SoC FPGA with a 166~MHz ARM Cortex M3 processor (SmartFusion2 M2S060T, Microsemi), enabling reprogrammable on-board computations. The Cortex has the available resources for local interference mitigation, feature extraction, and classification. While the current system does not require all of these resources, they will be included in future applications of the user-generic Ear EEG. The digitized EEG data are packetized and transmitted to the base station by a 2.4~GHz BLE radio (nRF51822, Nordic Semiconductor). The radio protocol also allows uplink from the base station to send configuration commands, start and stop recording, and set the frontend input range, etc. The BLE interface enables bidirectional wireless communication up to 2~m from the user at a data rate of 2~Mbps.

When streaming all 64~channels simultaneously, the device consumes 46~mW, of which \textless1~mW is dissipated by the NMIC, 10~mW are dissipated by the radio, while the remaining power is consumed by the SoC FPGA and power management circuits. Powered by a 3.7~V, 550~mAh lithium polymer (LiPo) battery, the device can operate for \mytilde44~hours continuously, though any 3.5~--~6~V battery may be used. For example, a smaller 300~mAh battery to provide \mytilde24~hours of operation. Data are received by a wireless base station connected to a laptop running a Python GUI for configuring the device, cueing the user, and real-time data visualization.

\vspace{-5pt}
\section{Experimental Results}

\subsection{Experimental setups}

\begin{figure}[tp]
\centerline{\includegraphics[width=2.5in]{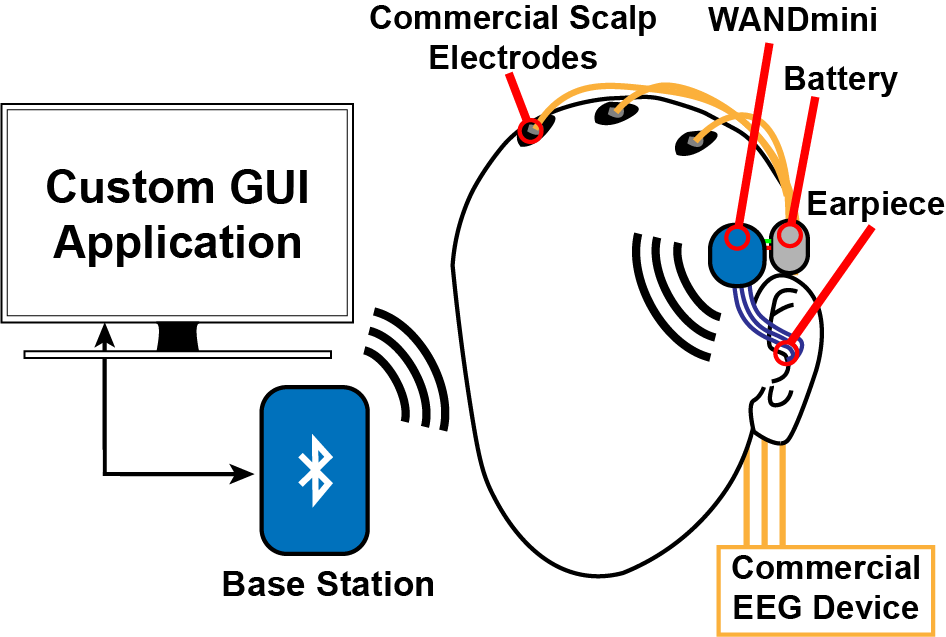}}
\caption{Experimental setup render. Head-worn WANDmini records EEG from the user-generic earpiece. All recorded data is sent to a base station via BLE. A custom GUI plots real-time data and cues subjects during experiments.}
\label{system_setup}
\vspace{-15pt}
\end{figure}

To verify the user-generic ear EEG system performance, electrophysiological measurements were performed to target three EEG paradigms, electrooculography (EOG), spontaneous EEG, and evoked EEG. The three electrophysiological signals selected to target these paradigms were eye blinks, alpha band modulation, and the auditory steady state response (ASSR), respectively. All measurements were simultaneously recorded by the ear EEG and a commercial EEG recorder (MPR ST+, Embletta) using wet Au cup electrodes on the scalp (Fig.~\ref{system_setup}). It is expected that scalp measurements will have higher SNR and greater signal amplitudes than in ear recordings due to their use of conductive electrode gels as well as the increased distance between electrodes. The larger, wet scalp electrodes have reduced ESIs and noise floors relative to dry electrodes, while recording across the entire scalp increases the recorded differential signal power\cite{Kappel2019}. Six wet electrodes were placed around the scalp at C3, C4, Cz, M1, M2, and the forehead as ground (according to the 10-20 system). All data recorded with the commercial EEG system was referenced against M1 and M2. The user-generic ear EEG was referenced and grounded separately. The ear EEG reference was the concha cymba electrode (Y, Fig.~\ref{manuf_photos}(d)) and was grounded through a wet electrode placed on the ipsilateral mastoid. Subject~1 used a “large” earpiece. Subjects 2 and 3 used “medium” earpieces. The user study was approved by UC Berkeley’s Institutional Review Board (CPHS protocol ID: 2018-09-11395). 

\vspace{-10pt}
\subsection{EOG: Eye Blinks}

Eye blinks, originating from both eye movements and EMG activity, generate large amplitude signals that are easily recorded with head-worn electrodes. While they are not cortical in origin, eye blinks have been increasingly incorporated into BCIs as control modes \cite{Schalk2004} as opposed to being treated as an artifact requiring removal \cite{Joyce2004}. Not only can eye blinks be used to select different options, but they can also be valuable features in drowsiness detection algorithms \cite{Nguyen2017}. To validate the user-generic ear EEG and characterize eye blinks as a possible BCI input, subjects were placed in front of a computer and told to blink when given an on-screen visual cue. The subjects would receive two types of cues: “hard blink” and “light blink”. All recordings were bandpass filtered from 0.05--50~Hz.

\begin{figure}[tp]
\centerline{\includegraphics[width=3in]{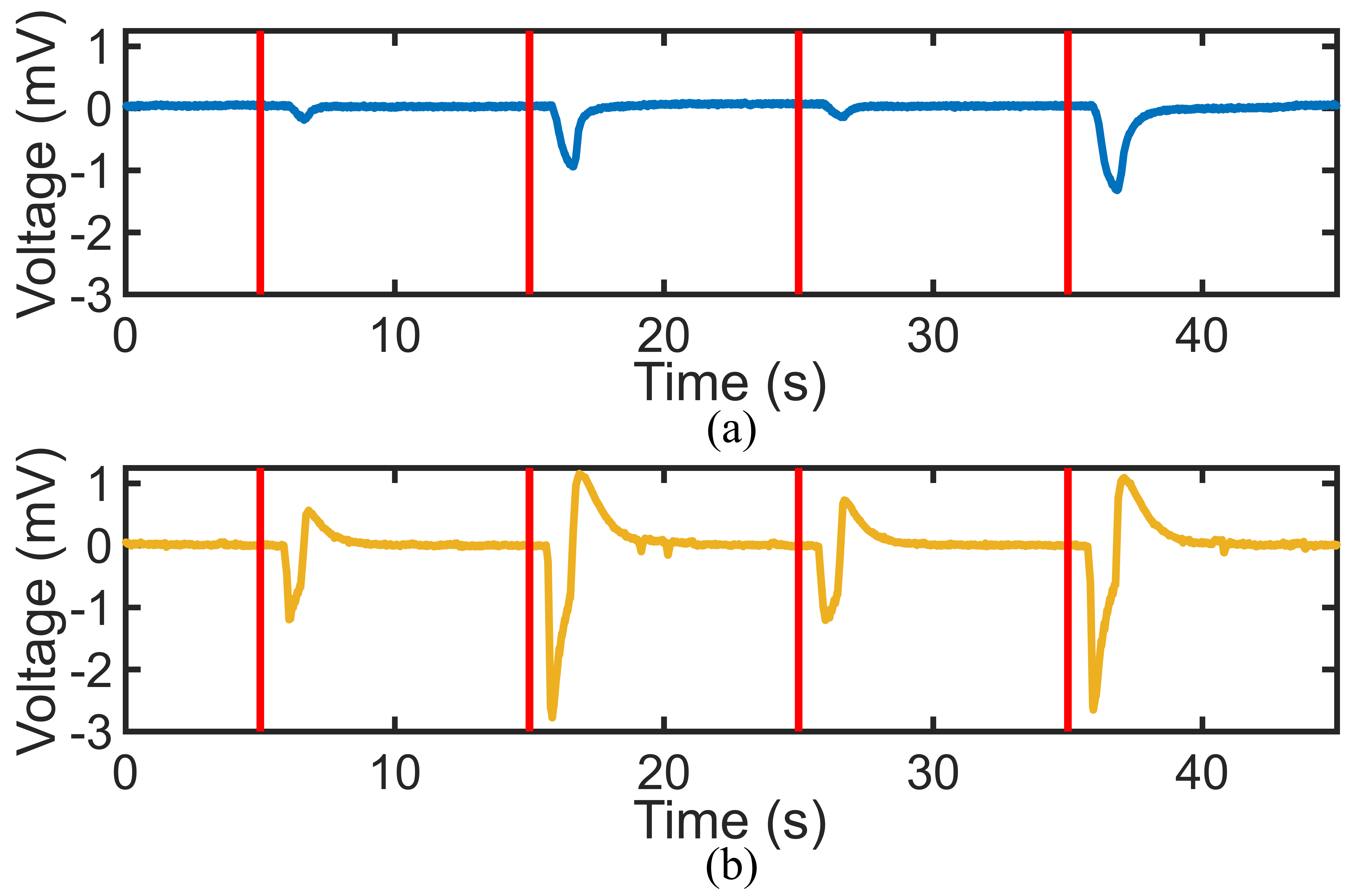}}
\vspace{-15pt}
\caption{Eye blinks recorded from subject 1 with the (a) user-generic ear EEG and (b) scalp EEG. Large blinks recorded with the user-generic ear EEG were 6x the amplitude of small blinks. Blinks recorded with the scalp EEG differ by 2.1x. Red lines mark visual cues.}
\label{blinks}
\vspace{-15pt}
\end{figure}

Fig.~\ref{blinks} shows eye blink amplitudes that alternate from \mytilde0.2~mV to \mytilde1.2~mV (factor of 6) in the ear EEG case, and \mytilde1.8~mV to \mytilde3.8~mV (factor of 2.1) when recorded with scalp EEG. The ear EEG recordings have an average baseline of 28~\SI{}{\micro V}rms while the scalp EEG recording’s average baseline is 16~\SI{}{\micro V}rms. Though the wet scalp electrodes recorded larger amplitude eye blinks (most likely due to the how scalp EEG electrodes are spaced further apart), all grades of eye blinks are clearly visible when recorded with the ear EEG.

\vspace{-10pt}
\subsection{Spontaneous EEG: Alpha Modulation}
Alpha rhythms are a spontaneous neural signal centered between 8--12~Hz that reflect a person's state of attention or relaxation. Originating from the occipital lobe, alpha rhythms are commonly used to benchmark EEG systems due to their prevalence and large amplitudes. As a result, the alpha band modulation has been extensively researched as a possible BCI feature \cite{Schalk2004}. Moreover, alpha rhythms have become useful signals for drowsiness detection and other neurofeedback applications\cite{Nguyen2017}. Alpha band power can be directly modulating by a person closing their eyes. Furthermore, user-to-user variability in the alpha wave amplitude is well documented \cite{Haegens2014}. To monitor the alpha wave with the ear EEG, subjects sat comfortably in a quiet room and were prompted to switch between two states every 30 s over the course of 2 min: an eyes open/focused state and an eyes closed/relaxed state.

\begin{table}[b]
\vspace{-10pt}
\setlength{\tabcolsep}{22pt}
\caption{ Individual and Grand Average Alpha Modulation Ratio Mean $\pm$ Std Dev (V\textsuperscript{2}/V\textsuperscript{2})}
\vspace{-15pt}
\begin{center}
\begin{tabular}{rcc} \toprule
& Ear EEG & Scalp EEG \\ \midrule
Subject 1 & $1.51 \pm 0.17$ & $5.29 \pm 1.5$ \\
Subject 2 & $2.44 \pm 0.62$ & $5.71 \pm 1.71$ \\
Subject 3 & $1.47 \pm 0.54$ & $3.06 \pm 1.53$ \\ \midrule
Average & $2.17 \pm 0.69$ & $5.54 \pm 1.84$ \\ \bottomrule
\end{tabular}
\label{tb:alpha}
\end{center}
\vspace{-10pt}
\end{table}

\begin{figure}[tp]
\centerline{\includegraphics[width=3.25in]{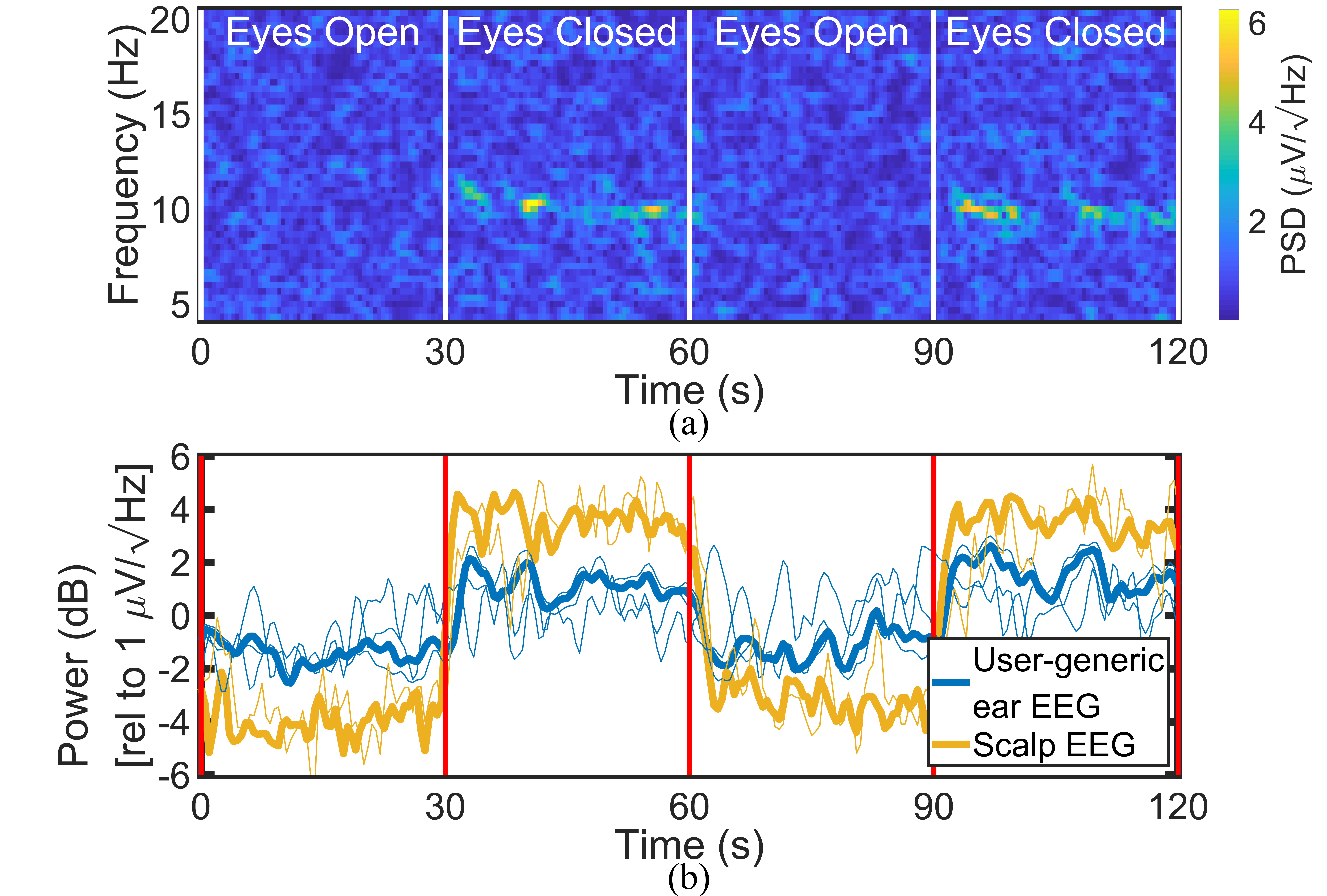}}
\vspace{-13pt}
\caption{(a) Time-frequency spectrogram of alpha modulation recorded with the user-generic ear EEG on subject 2. Alpha (8--12~Hz) power increased by a factor of 3.4 in the eyes-closed state. (b) Individual averages (light traces) and grand average (bold trace) of mean alpha power for three subjects and 44 total trials recorded with user-generic ear EEG and scalp EEG.}
\label{alpha}
\vspace{-15pt}
\end{figure}

The spectrogram in Fig.~\ref{alpha}(a) shows a representative example of a single user’s alpha modulation recorded using the user-generic ear EEG. Fig.~\ref{alpha}(b) shows the grand average of mean alpha (8--12 Hz) power from all three subjects. Mean alpha modulation $R_\up{AM}$ is defined as the ratio of mean alpha power from eyes closed to eyes open:
\begin{equation}\label{eq:alpha}
    R_\up{AM} = \frac{P_\up{avg}\lp\text{Alpha Band}_\text{Eyes Closed}\rp}{P_\up{avg}\lp\text{Alpha Band}_\text{Eyes Open}\rp}.
\end{equation}
The grand average alpha modulation (standard deviation) was 2.17~($\pm$ 0.69)~(V\textsuperscript{2}/V\textsuperscript{2}) for all user-generic ear EEG subjects and 5.54~($\pm$ 1.84)~(V\textsuperscript{2}/V\textsuperscript{2}) for wet scalp EEG. Specific subject alpha modulation ratios are shown in Table~\ref{tb:alpha} and in Fig.~\ref{alpha}(b). As expected, the scalp EEG recorded modulation is greater \cite{Kappel2019}, but the signal maintains sufficient SNR for user-generic ear EEG detection.

\vspace{-10pt}
\subsection{Evoked EEG: Auditory Steady-State Response}

\begin{figure}[tp]
\centerline{\includegraphics[width=3.25in]{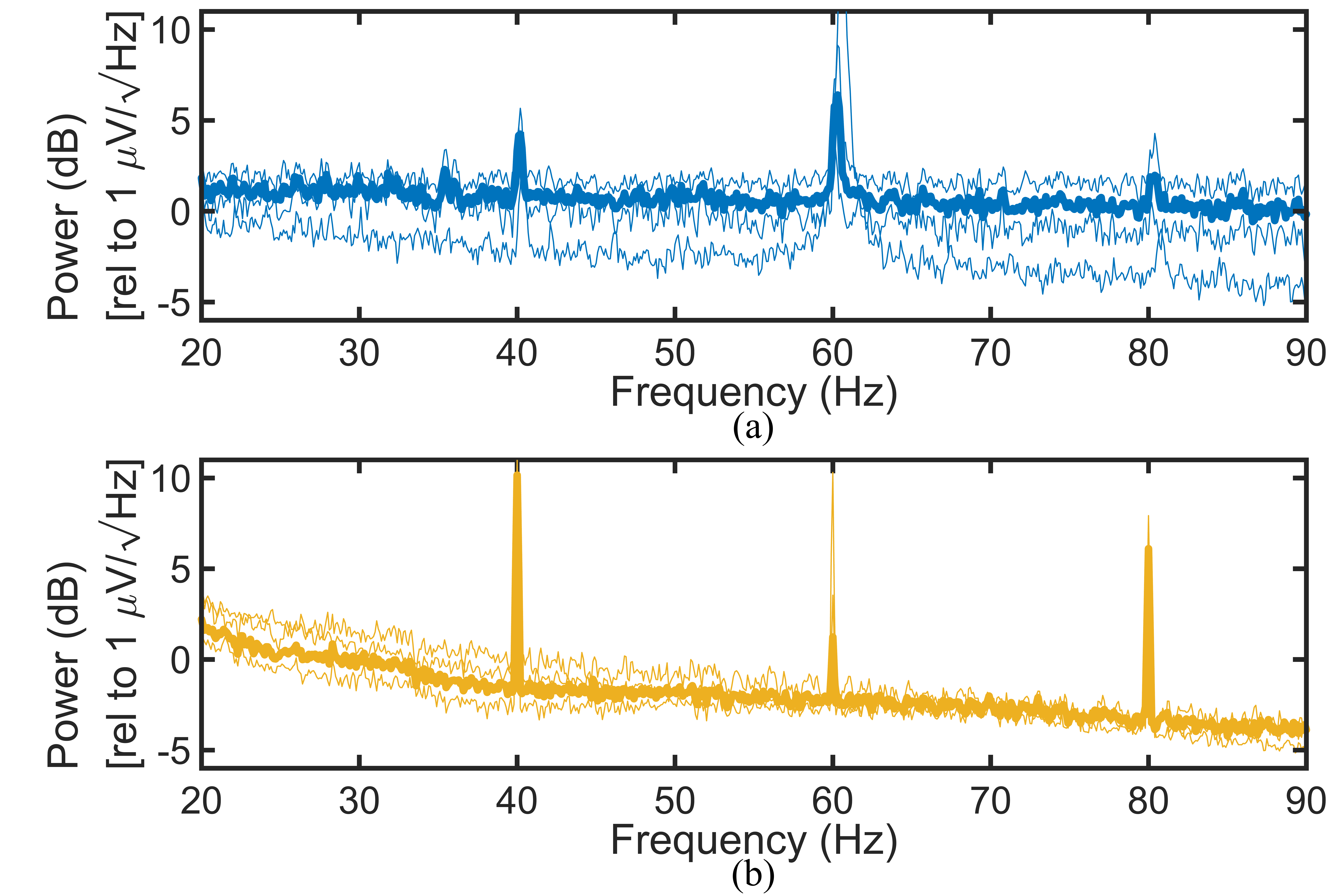}}
\vspace{-10pt}
\caption{Auditory evoked potentials (ASSR for 40~Hz click stimulus) grand and individual averages. (a) Grand average PSD of ASSR for three subjects and 41 total trials for user-generic ear EEG showing a mean SNR of \mytilde5.94~dB; (b) scalp EEG showing a mean SNR of \mytilde10.5~dB.}
\label{assr}
\vspace{-10pt}
\end{figure}

To confirm the ability to record evoked potentials with the user-generic ear EEG, ASSR, a neural response centered at the same frequency of a low frequency auditory stimulus, was targeted. Originating from the auditory cortex, ASSR can be evoked with different stimulus patterns, such as 40~Hz clicks or a 40~Hz amplitude-modulated 1~kHz tone \cite{Picton2003}. In addition, recent work has shown that ASSR is a powerful tool for performing automatic hearing threshold estimation in infants \cite{Rance2005} and determining binary choice via left-vs-right ear focus \cite{Lee2019}. To perform an ASSR experiment, each subject sat in a quiet room and listened to 40~Hz clicks sampled at 10~kHz played through desktop speaker for 100~s. Fig.~\ref{assr}(a) and (b) show the grand averages of data recorded with the user-generic ear EEG and scalp EEG, respectively. All power spectral densities show a clear neural response at both 40~Hz and the second harmonic at 80~Hz without any time-domain averaging.

The mean SNR was \mytilde5.94~dB for user-generic ear EEG and \mytilde10.5~dB for scalp EEG. Note that inter-user variability exists between users due to differences in their hearing ability, thus the user specific mean SNRs are shown in Table~\ref{tb:assr} (and are also represented in Fig.~\ref{assr}). Mean SNR was calculated as the ratio of the power at 40~Hz to the mean power from 35--45~Hz, excluding 40~Hz \cite{Kappel2019}: 

\begin{equation}
    \text{40 Hz SNR} = \frac{P\lp\text{40 Hz}\rp}{P_\up{avg}\lp\text{35 -- 45 Hz}\rp^\ast}.
\end{equation}
\[^\ast \text{excluding 40 Hz}\]
\vspace{-3pt}
While the SNR is lower for user-generic ear EEG (due to higher noise floor and shorter electrode distance, which reduces signal amplitude \cite{Kappel2019}), the SNR maintains sufficient SNR for user-generic ear EEG detection.

\begin{table}[h]
\setlength{\tabcolsep}{22pt}
\caption{Individual and Grand Average ASSR 40~Hz SNR $\pm$ Std Dev (dB)}
\vspace{-15pt}
\begin{center}
\begin{tabular}{rcc} \toprule
& Ear EEG & Scalp EEG \\ \midrule
Subject 1 & $5.64 \pm 1.39$ & $13.79 \pm 6.08$ \\ 
Subject 2 & $5.40 \pm 0.99$ & $7.36 \pm 2.71$ \\
Subject 3 & $6.80 \pm 2.02$ & $8.51 \pm 2.42$ \\ \midrule
Average & $5.94 \pm 1.70$ & $10.48 \pm 5.23$ \\ \bottomrule
\end{tabular}
\label{tb:assr}
\end{center}
\vspace{-10pt}
\end{table}

\begin{table*}[tbp]
\begin{center}
\setlength{\tabcolsep}{7.3pt}
\caption{Table Comparing Recent In-ear EEG Systems}
\begin{tabular}{rcccccccc}\toprule
& \cite{Kidmose2012} & \cite{Kidmose2013} & \cite{HoonLee2014} & \cite{Goverdovsky2016} & \cite{Dong2016} & \cite{Lee2019} & \cite{Kappel2019} & This work \\ \midrule 
Electrode Material & Wet Ag & Wet Ag & \begin{tabular}[c]{@{}c@{}}Dry\\ CNT/PDMS\end{tabular} & \begin{tabular}[c]{@{}c@{}}Wet Ag-\\ coated nylon\end{tabular} & \begin{tabular}[c]{@{}c@{}}Dry silvered\\ glass silicone\end{tabular} & Dry & Dry IrO2 & Dry Ag \\  \midrule 

Dry Electrodes       & -- & -- & 1 & -- & 1 & 3 & 6 & 6 \\   \midrule
Reference            & Wet & Wet & Wet & Wet & Wet & Dry & Dry & Dry \\ \midrule 
Electrode Area       & -- & 15 mm\textsuperscript{2} & -- & 40 mm\textsuperscript{2} & -- & -- & 9.6 mm\textsuperscript{2} & 60 mm\textsuperscript{2} \\  \midrule 
Impedance at 50 Hz   & -- & -- & 50 k\(\Omega\) & -- & -- & -- & 1.1 M\(\Omega\) & 377 k\(\Omega\) \\   \midrule
Wireless             & No & No & No & No & No & Yes (100 kbps) & No & Yes (2 Mbps) \\  \midrule 
Earpiece Style       & Custom & Generic & Generic & Generic & Generic & Custom & Custom & Generic \\  \midrule 
Scalable             & No & No & No & No & Yes & No & No & Yes \\ \midrule
Mean Alpha Modulation & -- & -- & -- & -- & 1.5* & -- & 1.2 & 2.1 \\ \bottomrule
\end{tabular}
\label{tb:comp_table}
{\newline \raggedright *estimated **sense electrode area only \par}
\end{center}
\vspace{-18pt}
\end{table*}

\vspace{-10pt}
\section{Summary}
Traditional scalp EEG systems have demonstrated BCIs that enable prosthetic control and choice selection, but have been too cumbersome (tedious set-up, increased risk of skin lesion formation, susceptible to motion artifacts, etc.) to be used in a daily and public setting. Recording EEG from inside the ear canal with dry-electrodes has introduced the possibility of integrating BCIs into commonly used earbuds. These envisioned neural wearables would improve ease-of-use and discreetness by incorporating the wireless recording module into the earpiece itself. Such a user-generic neural recording wearable would require a generalized earpiece employing dry electrodes that maintain electrical characteristics over long time periods. The wireless recording electronics should be designed for dry electrode noise and interference properties and be capable of ambulatory recordings.

This work presents a discreet ear EEG recording system with a user-generic earpiece and wireless neural recording module. To promote deployment across large test groups, a low-cost, scalable manufacturing process with commonly-used materials was developed alongside a small form factor, wireless neural recording module. ESI, EDO, and noise characterizations were performed over the course of six months and showed little to no degradation in the electrode, thus motivating the earpiece's long-term usability. A user study across three subjects was performed in which eye blinks, alpha modulation, and ASSR were recorded without the need for time-domain averaging to motivate future BCI applications such as blink activation of voice assistants, stress detection, and automated hearing threshold detection.

To the best of the authors' knowledge, this is the only work that supports multi-channel wireless user-generic ear EEG recording across multiple users with the same earpiece design (Table~\ref{tb:comp_table}). The presented earpiece design leverages greater electrode areas to reduce ESI relative to \cite{Kappel2019}. In addition, it is the only user-generic earpiece that does not require exotic materials or the use of wet reference electrodes placed outside the ear (\cite{Lee2019} \cite{Dong2016}). Mean alpha modulation was used to compare the user-generic ear EEG platform's sensitivity with a state-of-the-art dry electrode in-ear EEG system. Though user-to-user variability may play a role, the user-generic ear EEG platform recorded a mean alpha modulation of 2.17 while the best reported prior art \cite{Kappel2019} recorded a mean alpha modulation of 1.2 (Table~\ref{tb:comp_table}).
\vspace{-10pt}
\section*{Acknowledgment}
\vspace{-4pt}
The authors thank the sponsors of the Berkeley Wireless Research Center, the Ford University Research Program, Cortera Neurotechnologies and Formtech. Thanks to Prof. Robert Knight for technical discussion.


\vspace{-11pt}
\bibliographystyle{IEEEtran}
\bibliography{bibliography}

%
%






\end{document}